\documentclass{kapproc}
\normallatexbib

\begin{document}

\articletitle[Boundary O(N) Nonlinear Sigma Model]
{Integrable Boundary Conditions \\
for the O(N) Nonlinear $\sigma$ Model}

\author{Marco Moriconi}

\affil{Newman Laboratory\\
Cornell University, Ithaca, NY 14853\\
U.S.A.\footnote{Partial funding provided by NSF}}

\email{moriconi@mail.lns.cornell.edu}

\begin{abstract}
We discuss the new integrable boundary conditions for the $O(N)$ nonlinear
$\sigma$ model and related solutions of the boundary Yang-Baxter equation,
which were presented in our previous paper hep-th/0108039.

\end{abstract}


\section{Introduction}
Two-dimensional nonlinear $\sigma$ (nl$\sigma$) models have been the subject of
intense study during the past few years, since they may be used as toy
models for the study of higher dimensional non-abelian gauge theories
(Yang-Mills), they arise in several condensed matter and statistical
mechanics problems, and there are powerful mathematical methods in 2d
that allow one to have a deeper understanding of their structure. On
top of that, they display a host of theoretical phenomena, such as
asymptotic-freedom, dynamical mass generation, and $1/N$-expansions.

A natural generalization, both from the theoretical and experimental
point of view, of a given 2d integrable model, is to consider it on the
half-line \cite{GZ}. This type of reduction arises in several problems, for
example, when considering the radial part of the Schroedinger
equation for a radial potential, in the study of quantum impurities,
such as the Kondo effect, and in open string theory.

In this note we summarize the results obtained in \cite{M}, where we
have found new integrable boundary conditions and related solutions of
the boundary Yang-Baxter equation (bYBe) for the $O(N)$ nl$\sigma$ model. The
following discussion is informal and intended to a general audience,
we refer to \cite{M} for a more complete discussion.

\section{The $O(N)$ Nonlinear Sigma Model}

The lagrangian of the $O(N)$ nl$\sigma$ model \cite{P} is
\begin{equation}
{\cal L}_{nl \sigma}=\frac{1}{2g_0^2}(\partial{\vec {\bf n}})^2
\end{equation}
where the field ${\vec {\bf n}}=(n_1,n_2, \ldots, n_N)$ satisfies
${\vec {\bf n}}^2=1$, and $g_0$ is a coupling constant. This
constraint may be introduced in the lagrangian through a lagrangian
multiplier $\lambda(x)$,
\begin{equation}
{\cal L}_{nl\sigma}=\frac{1}{2g_0^2}(\partial{\vec {\bf n}})^2+
\lambda(x)({\vec {\bf n}}^2-1) \label{nls}
\end{equation}
The equation of motion for the field ${\vec {\bf n}}$ is then easily
derived from (\ref{nls}) and, in light-cone coordinates $x_{\pm}=(x_0 \pm
x_1)/2$, it reads
\begin{equation}
\partial_+\partial_-{\vec {\bf n}}=-{\vec {\bf n}} \,\,
(\partial_+{\vec {\bf n}} \cdot \partial_-{\vec {\bf n}})
\end{equation}
The exact $S$-matrix for te $O(N)$ nl$\sigma$ model was found by Zamolodchikov and 
Zamolodchikov in \cite{ZZ}. For a review on exact $S$-matrices see \cite{ZZ2}.

Since the $O(N)$ nl$\sigma$ model is classically conformally invariant (no
dimensionful quantities appear in the classical lagrangian
(\ref{nls})) the trace of the energy-momentum tensor, $T_{+-}$($=T_{-+})$,
vanishes. Upon quantization, conformal invariance is broken due to the
introduction of an energy scale, such as an energy cut-off, for
example. In any case we will see later that the classical conformal
invariance is crucial in establishing the quantum integrability of
this model. In light-cone coordinates the only non-vanisihing
components of the energy momentum tensor are $T_{++}=\partial_+{\vec
{\bf n}} \cdot \partial_+{\vec {\bf n}}$ and $T_{--}=\partial_-{\vec {\bf n}}
\cdot \partial_-{\vec {\bf n}}$.  This means that energy-momentum
conservation reads
\begin{equation}
\partial_-T_{++}=0 \label{t++}
\end{equation}
and an analogue equation with $+ \leftrightarrow -$. Classically it is
easy to see that (\ref{t++}) implies $\partial_-(T_{++})^n=0$ for any
integer $n$.  Upon quantization this is no longer true since we have
now a composite operator and one must be careful in defining the
product of operators at the same space-time point. This means that the
right-hand side of (\ref{t++}) will no longer be zero, or in other
words, we have a quantum anomaly. Even though it is a hard problem to
compute the exact form, with all numerical coefficients, of the anomaly
in (\ref{t++}), we can greatly fix its form, by using dimensional
analysis and group theory.  Let us look at what happens in the case
$n=2$. In this case the rhs of (\ref{t++}) has dimension 5, Lorentz weight 3
\footnote{If a quantity $\phi$ transforms as $\phi \rightarrow
\exp(p\alpha)\phi$ under a Lorentz boost, that is, the rapidity
variable $\theta \rightarrow \theta + \alpha$, we say that $\phi$ has
Lorentz weight $p$.}, and it is a scalar under the global $O(N)$.
This means that, whatever the anomaly is, it is a local operator that
satisfies these three requirements. One can then proceed and make a
list of the possible operators that contribute to the anomaly. The rhs
of (\ref{t++}) will be, finally, a linear combination of these operators, the
hard work being to have to compute the coefficients of this linear
combination. In the case of $\partial_- (T_{++})^2=0$ a wonderful
thing happens: all the operators that can contribute to the anomaly
can be rewritten as a total derivative, with the help of the equation
of motion and the constraint ${\vec{\bf n}} \cdot {\vec{\bf
n}}=1$. This is the so-called Goldschmidt-Witten argument \cite{GW}
(see also \cite{P}, for an earlier version). For the complete list of
these operators, we refer to \cite{GW}. By using this quantum
conservation law the integrability of the $O(N)$ nl$\sigma$ model
is established.

When considering a field theory on the half-line, bulk conservation laws may
be broken, as it is clear in the case of linear momentum conservation 
(the boundary breaks translation symmetry). Therefore, one needs to impose
suitable boundary conditions that will ensure that a given bulk conservation
law will still hold after the introduction of the boundary. In 
equations, if we have a bulk conservation law of spin (Lorentz weight) $s$
\begin{equation}
\partial_+J_-^{(s+1)}=\partial_-R_+^{(s-1)} \qquad {\rm{and}} \qquad 
\partial_-J_+^{(s+1)}=\partial_+R_-^{(s-1)}
\end{equation}
it follows that the charges
\begin{equation}
Q_{\pm}=\int_{-\infty}^{+\infty} dx_1 \; (J_{\pm}^{(s+1)}- R_{\mp}^{(s-1)} ) 
\end{equation}
are conserved. After the introduction of a boundary, only (possibly) a 
combination of these charges will be conserved. The charge is
\begin{equation}
{\widetilde Q}=
\int_{-\infty}^0 dx_1 \; (J_{+}^{(s+1)}- R_{-}^{(s-1)}+
J_{-}^{(s+1)}- R_{+}^{(s-1)})+\Sigma(t)
\end{equation}
where $\Sigma(t)$ is a local operator satisfying
\begin{equation}
\left. J_{-}^{(s+1)} -J_{+}^{(s+1)} +
R_{-}^{(s-1)} -R_{+}^{(s-1)} \right|_{x=0} = \frac{d}{dt}\Sigma(t)
\ , \ \label{condition}
\end{equation}
This is precisely where we have to impose the boundary conditions in order to
have a non-trivial conserved charge in the presence of a boundary.

In \cite{MDM} we have shown that if we impose Neumann boundary
conditions, $\partial_1 n_i|_{x=0}=0$, to $k$ field components, and
Dirichlet boundary conditions, $\partial_0 n_i|_{x=0}=0$, to the
remaining $N-k$ components, the condition (\ref{condition}) is
satisfied and we have integrable boundary conditions.  Note that there
is no free parameter (coupling constant) in this case.  These are
diagonal boundary conditions, in the sense that scattering off the
boundary does not change the $O(N)$ index of the incoming particle.
These boundary conditions break the bulk symmetry at the boundary to
$O(k) \times O(N-k)$. It can be shown, by using the bYBe, that these
are the only possible diagonal integrable boundary
conditions. Therefore, if we are looking for new integrable boundary
conditions with free parameters, we have to necessarily look for
non-diagonal boundary conditions, that is, boundary conditions that
change the $O(N)$ index (flavor) of the incoming particle through
scattering off the boundary.

In the following we will take a slightly different point of view from
\cite{M}, where we used a simple two free bosons model coupled at the
boundary as a guide to the possible boundary conditions for the $O(N)$
nl$\sigma$ model.

In \cite{M} we have found new integrable boundary conditions that
break the bulk $O(N)$ symmetry to $O(2) \times O(N-2)$ at the
boundary, and which depend on one free-parameter $g$. The reason for
this symmetry at the boundary is the following. Free boundary
conditions (Neumann) have $O(N)$ boundary symmetry. The diagonal
boundary conditions we found previously, break the boundary symmetry
to $O(k) \times O(N-k)$. We are looking now for non-diagonal boundary
conditions with a free parameter, and we may assume that in certain
limiting cases, such as taking the coupling constant to $0$ or
$\infty$, we should reduce to a diagonal case. Therefore, if we insist
that once the boundary symmetry is broken we do not have any point in
the integrable flow where it is enhanced, we should look for
non-diagonal boundary conditions that are $O(k) \times O(N-k)$
symmetric.  By writing the generic boundary condition as
\begin{equation}
\partial_1 n_i|_{x=0} =  M_{ij} \,\, \partial_0 n_j |_{x=0}
\end{equation}
where the indices $i$ and $j$ run through a subset of
$\{1,2,\ldots,N\}$ (the ``nondiagonal subset'', which could be taken
to be the first $k$ indices, for example). Under an orthogonal
transformation the fields transform as $n_i \rightarrow {\tilde n_a} =
O_{ai} { n_i}$, where $O$ is a $k \times k$ orthogonal matrix. This
means that $M_{ij} \rightarrow {\tilde M_{ab}}= O_{ai}M_{ij}O_{bj}$,
and if we require the boundary conditions to be $O(k) \times O(N-k)$
symmetric, we should have $O M O^t=M$. The only case where we can
impose this condition for a non-diagonal matrix if when $k=2$, since
$O(2)$ is abelian. This fixes the matrix $M$ to be of the form $M=g_1
I + i \, g_2 \sigma_2$, where $I$ is the identity matrix and
$\sigma_2$ is a Pauli matrix. By inspecting the spin-$4$
Goldschimdt-Witten charge described above we see that if we take
$g_1=0$ and $g_2=g$ arbitrary, the following boundary condition is
integrable:
\begin{equation}
\partial_1 n_1|_{x=0} =  g \,\, \partial_0 n_2 |_{x=0} \qquad {\rm{and}} \qquad
\partial_1 n_2|_{x=0} = -g \,\, \partial_0 n_1 |_{x=0} \label{bc}
\end{equation}
where we picked the first two components of the ${\vec {\bf n}}$ field
without any loss of generality. The remaining field components satisfy
Dirichlet boundary condition \footnote{This choice of boundary
condition for the remaining field components will become clear when
we discuss the boundary Yang-Baxter equation.}. In a different form,
this boundary condition had been studied by Corrigan and Sheng at the
classical level in \cite{CS}, for the $O(3)$ nl$\sigma$ model.

The non-diagonal boundary conditions in (\ref{bc}) can be derived from 
the boundary lagrangian ${\cal L}_b=\frac{1}{2}M_{ij}n_i {\dot n}_j$, which
shows that $M_{ij}$ should be anti-symmetric.

By taking $g \rightarrow 0$ we have diagonal boundary conditions,
where 2 field components satisfy Neumann and the remaining Dirichlet,
and by taking $g \rightarrow \infty$ we recover a diagonal case again,
with all field components satisfying Dirichlet boundary conditions.
Therefore we have an integrable flow between diagonal boundary
conditions, from $O(N)$, corresponding to $g=\infty$, to $O(2) \times
O(N-2)$, corresponding to $g=0$.

\section{The Reflection Matrix}

When one tries to find an exact $S$-matrix for a given integrable
field theory, the use of the bulk symmetries plays a crucial role,
making it much easier to solve the Yang-Baxter equation. This is why
we had to understand the symmetry of the boundary conditions before we
could go on and try to solve the bYBe.

For the purely diagonal case, the solutions of the bYBe have been
found in \cite{MDM1}. They are block diagonal, $O(k) \times O(N-k)$
symmetric, with diagonal elements $(R_1(\theta), \ldots, R_2(\theta)
\ldots)$, the first $k$ elements corresponding to Neumann, and
the remaining $N-k$ to Dirichlet. The bYBe fixes the ratio
$R_1(\theta)/R_2(\theta)$ to be
\begin{equation}
\frac{R_1(\theta)}{R_2(\theta)}=\frac{c-\theta}{c+\theta}
\end{equation}
with $c=-i\frac{\pi}{2}\frac{N-2k}{N-2}$. Note that there is an interesting
duality by taking $k \rightarrow N-k$, which takes $c \rightarrow -c$, and
therefore $R_1(\theta) \leftrightarrow R_2(\theta)$.

For the boundary conditions (\ref{bc}), we start with the following ansatz
\begin{equation}
R=\left(\begin{array}{ccccc}
                    \phantom{-} A(\theta) & B(\theta) & 0 & 0 & \cdots 
\\
                    -B(\theta) & A(\theta) & 0 & 0 & \cdots \\
                    \phantom{-} 0 & 0 & R_0(\theta) & 0 & \cdots  \\
                    \phantom{-} 0 & 0 & 0 & R_0(\theta) & \cdots \\
                    \phantom{-} \vdots & \vdots &  \vdots  & \vdots & 
\ddots
        \end{array}\right) \ . \ \label{ansatz}
\end{equation}
This means that the first two particles can scatter onto each other with
amplitude $\pm B(\theta)$, or onto themselves with amplitude $A(\theta)$.
The diagonal elements correspond to the particles scattering diagonally with
Dirichlet boundary conditions, with amplitude $R_0(\theta)$. Thinking in
terms of the boundary lagrangian for the non-diagonally scattering particles,
we see that the off-diagonal amplitudes should have opposite signs.

We can use the bYBe now, in order to compute the functions
$A(\theta)$, $B(\theta)$, and $R_0(\theta)$. In the following we will
quote the differential equations for $X(\theta)=A(\theta)/R_0(\theta)$
and $Y(\theta)=B(\theta)/R_0(\theta)$, obtained from the bYBe, by
taking the limit where the two rapidities are equal.

The process $|A_1(\theta_1)A_i(\theta_2)\rangle \rightarrow
|A_i(-\theta_1)A_1(-\theta_2)\rangle$ \footnote{The $\{A_i(\theta)\}$
are the usual Faddeev-Zamolodchikov operators.}, where $i$ is any of
the diagonally scattering particles, gives
\begin{equation}
\frac{d}{d\theta}X(\theta)=\frac{X^2(\theta)-Y^2(\theta)-1}{2\theta} \ . \
\end{equation}
The process $|A_1(\theta_1)A_i(\theta_2)\rangle\rightarrow
|A_i(-\theta_1)A_2(-\theta_2)\rangle$ gives
\begin{equation}
\frac{d}{d\theta}Y(\theta)=\frac{X(\theta)Y(\theta)}{\theta} \ . \
\end{equation}
These two equations can be easily solved by the introduction of the
auxiliary functions $Z_{\pm}(\theta)=X(\theta) \pm i Y(\theta)$. We obtain
\begin{equation}
X(\theta)=\frac{1}{2}\left(\frac{c-\theta}{c+\theta}+
\frac{c'-\theta}{c'+\theta}\right) \qquad {\rm{and}} \qquad
Y(\theta)=\frac{1}{2i}\left(\frac{c-\theta}{c+\theta}-
\frac{c'-\theta}{c'+\theta}\right)
\end{equation}
where $c$ and $c'$ are constants to be determined. Since we have only
one free parameter at the boundary, we should find one
equation relating $c$ and $c'$. This is acomplished by the bYBe
corresponding to the process $|A_1(\theta_1)A_1(\theta_2)\rangle
\rightarrow |A_1(-\theta_1)A_2(-\theta_2)\rangle$, from which we
obtain
\begin{equation}
c+c'=-i\pi\frac{N-4}{N-2}
\end{equation}
We have verified that with this constraint, all the other bYBe's are
satisfied.  Once the ratios $X(\theta)$ and $Y(\theta)$ have been
fixed, all that is left to do is to compute the overall factor for the
reflection matrix, which can be done with the use of boundary
unitarity and boundary crossing-symmetry, and a minimality hypothesis
for the pole structure of the reflection matrix.  We refer to \cite{M}
for the explicit results.

Note that if $c=c'$ the off-diagonal amplitudes $\pm Y(\theta)$
vanish, and we recover a diagonal scattering problem. The other
instance where $Y(\theta)$ vanishes is when $|c|, |c'| \rightarrow
\infty$.  In the first case the ratio
$X(\theta)=\frac{c-\theta}{c+\theta}$ with
$c=-i\frac{\pi}{2}\frac{N-4}{N-2}$, which corresponds precisely to the
case where the first two field components satisfy Neumann boundary
conditions and the remaining $N-2$ Dirichlet. This is the reason why
we chose the diagonally scattering field components to satisfy
Dirichlet boundary conditions. In this case the solution for the
reflection matrix is $O(2) \times O(N-2)$ symmetric. By looking at the
explicit form of the boundary conditions (\ref{bc}), we see that this
corresponds to $g=0$.  In the second case $X(\theta)=1$, which means
that the reflection matrix is proportional to the identity, and
therefore $O(N)$ symmetric. This corresponds to taking $g \rightarrow
\infty$, and therefore, to all components satisfying Dirichlet
boundary conditions.

We can introduce the following conveninent parametrization:
$c=-i\frac{\pi}{2}\frac{N-4}{N-2}+\xi(g)$ and
$c'=-i\frac{\pi}{2}\frac{N-4}{N-2}-\xi(g)$, where $\xi(g)$ is an
unknown function of the boundary coupling constant. The two cases described
in the preceding paragraph correspond to $\xi(0)=0$ and $\xi(g \rightarrow 
\infty) \rightarrow \infty$. This establishes an integrable flow between
different diagonal boundary conditions.

One could be tempted at trying a generalization of the ansatz
(\ref{ansatz}), with more than one non-diagonal block, corresponding
to more than one pair of particles being coupled at the boundary. By
using the bYBe it can be shown that there are no solutions of this
type \cite{M}.

In \cite{M} we found other solutions to the bYBe for the $O(2N)$ nl$\sigma$
model, but were not able to link them to any boundary
conditions. Another special case is the $O(2)$ nl$\sigma$ model. Naively
one could be lead to think that the $O(2)$ nl$\sigma$ model is equivalent to
a massless free boson, through a mapping $(n_1,n_2) \rightarrow
(\cos(\theta),\sin(\theta))$, but this is not the case, and after a
more careful analysis, it can be shown that the $O(2)$ nl$\sigma$ model is
equivalent to the sine-Gordon model at $\beta^2=8\pi$, which
describes the Kosterlitz-Thouless point of the classical $XY$
model. The solution we found depends on three parameters, instead of
two as in the boundary conditions found by Ghoshal and Zamolodchikov
in {\cite{GZ}. The resolution of this discrepancy is that we are looking at
the sine-Gordon model at a special value of the coupling constant,
and as already remarked in \cite{GZ}, at these special points there are more
solutions than the ones found for the general case.

\section{conclusions}

We have found new integrable boundary conditions for the $O(N)$ nl$\sigma$
model, which depend on one free parameter $g$. These bounday
conditions break the bulk $O(N)$ symmetry to $O(2) \times O(N-2)$, and
by taking the limits $g \rightarrow 0$ and $g \rightarrow \infty$ we recover
diagonal solutions studied previously. This establishes an integrable
flow between two different sets of boundary conditions.

Recently Mackay and Short \cite{MS} have studied the principal chiral model
with a bounday, and found an interesting relationship between their
boundary conditions and the theory of symmetric spaces. Their
solutions, though, are quite different from ours, and some work should
be done in trying to clarify their relationship.

As natural follow-up problems, one should try to find explicitly the
function $\xi(g)$ in the reflection matrices, and to study the boundary
thermodynamic Bethe ansatz equations.

An interesting direction to pursue would be to extend these results to
the $SO(N)$ Gross-Neveu (GN) model. Since the $S$-matrix for the
elementary fermions of the GN model is equivalent to the one for the
$O(N)$ nl$\sigma$ model, up to a CDD factor, we certainly can find solutions
of the bYBe of the form (\ref{ansatz}) for the GN model too.

\begin{acknowledgments}
I would like to thank the organizers of the NATO Advanced
Research Workshop on ``Statistical Field Theories'', G. Mussardo
and A. Cappelli, for organizing such a stimulating workshop and for
the invitation to present these results there. I would also like to thank
the hospitality of the Abdus Salam ICTP and SISSA, where part of this 
work was done, and to T. Becher, A. Petrov, and V. Sahakian, for several
dicussions.
\end{acknowledgments}


\begin{chapthebibliography}{1}

\bibitem{GZ} S. Ghoshal and A. Zamolodchikov, 
{\em ``Boundary $S$-Matrix and Boundary State in Two-Dimensional Integrable Quantum 
Field Theory''},
Int.\ J.\ Mod.\ Phys.\ A {\bf 9}, 3841-3886 (1994), 
[hep-th/9306002].

\bibitem{M} 
M.~Moriconi, 
{\em ``Integrable boundary
conditions and reflection matrices for the O(N) nonlinear sigma
model''}, to appear in Nucl. Phys. {\bf B},
[hep-th/0108039].

\bibitem{P} 
A.~M.~Polyakov,
{\em ``Hidden Symmetry Of The Two-Dimensional Chiral Fields''},
Phys.\ Lett.\ B {\bf 72}, 224 (1977).

\bibitem{ZZ}
A.~B.~Zamolodchikov and A.~B.~Zamolodchikov,
{\em ``Relativistic Factorized S Matrix In Two-Dimensions Having O(N) Isotopic Symmetry''},
Nucl.\ Phys.\ B {\bf 133}, 525-535 (1978) [JETP Lett.\  {\bf 26}, 457 (1978)].

\bibitem{ZZ2}
A.~Zamolodchikov and A.~Zamolodchikov,
{\em ``Factorized S-Matrices In Two Dimensions As The Exact Solutions 
Of  Certain Relativistic Quantum Field Models''},
Annals Phys.\  {\bf 120}, 253-291 (1979).

\bibitem{GW} 
Y.~Y.~Goldschmidt and E.~Witten,
{\em ``Conservation Laws In Some Two-Dimensional Models''},
Phys.\ Lett.\ B {\bf 91}, 392-396 (1980).

\bibitem{MDM}
M.~Moriconi and A.~De Martino,
{\em ``Quantum integrability of certain boundary conditions''},
Phys.\ Lett.\ B {\bf 447}, 292-297 (1999)
[hep-th/9809178].

\bibitem{CS}
E.~Corrigan and Z.~Sheng,
{\em ``Classical integrability of the O(N) nonlinear sigma model on a  
half-line''}
Int.\ J.\ Mod.\ Phys.\ A {\bf 12}, 2825-2834 (1997)
[hep-th/9612150].

\bibitem{MDM1}
A.~De Martino and M.~Moriconi,
{\em ``Boundary S-matrix for the Gross-Neveu model''},
Phys.\ Lett.\ B {\bf 451}, 354-364 (1999)
[hep-th/9812009].

\bibitem{MS}
N.~J.~MacKay and B.~J.~Short,
{\em ``Boundary scattering, symmetric spaces and the principal chiral model on  the half-line''},
[hep-th/0104212].

\end{chapthebibliography}
\end{document}